# Investigation of Shock Wave Interactions involving Stationary and Moving Wedges

Pradeep Kumar Seshadri and Ashoke De[*]

*Department of Aerospace Engineering, Indian Institute of Technology Kanpur, Kanpur, 208016, India.*

*Corresponding Author: ashoke@iitk.ac.in

## Abstract

The present study investigates the shock wave interactions involving stationary and moving wedges using a sharp interface immersed boundary method combined with a fifth-order weighted essentially non-oscillatory (WENO) scheme. Inspired by Schardin's problem, which involves moving shock interaction with a finite triangular wedge, we study influences of incident shock Mach number and corner angle on the resulting flow physics in both stationary and moving conditions. The present study involves three incident shock Mach numbers (1.3, 1.9, 2.5) and three corner angles (60°, 90°, 120°), while its impact on the vorticity production is investigated using vorticity transport equation, circulation, and rate of circulation production. Further, the results yield that the generation of the vorticity due to the viscous effects are quite dominant compared to baroclinic or compressibility effects. The moving cases presented involve shock driven wedge problem. The fluid and wedge structure dynamics are coupled using the Newtonian equation. These shock driven wedge cases show that wedge acceleration due to the shock results in a change in reflected wave configuration from Single Mach Reflection (SMR) to Double Mach Reflection (DMR). The intermediary state between them, the Transition Mach Reflection (TMR), is also observed in the process. The effect of shock Mach number and corner angle on Triple Point (TP) trajectory, as well as on the drag coefficient, is analyzed in this study.

## I. Introduction

Shock interactions with solid bodies generate complex flow fields that involve several phenomena such as reflection, diffraction, shock-boundary layer interaction, shock-vortex interaction, shock-shock interaction, and so on. Developing deeper insights into these interactions can help in addressing issues in drug delivery[1], material processing[2], or understanding geophysical processes[3] apart from practical applications in aerospace systems. Since the early 1950s, several experimental and computational studies have sought to understand shock-body interactions leading to shock propagation, reflection, and diffraction patterns by studying moving shock interaction with simple two-dimensional wedge configurations involving concave and convex corners.

Shock reflection phenomena over an inclined surface have been extensively studied for the past several decades. Ben-dor[4] in his monograph had documented 13 different possible wave configurations in steady, pseudo-steady and unsteady flows and has mapped the transition boundary of various wave configurations as a function of moving shock Mach number $M_i$ (between 1.0 to 4.0) and wall inclination angle $\theta$ (between 0° to 60°). The shock reflection has also been studied over double wedge configurations[5-9] mainly to study the transition of a shock wave from one wave configuration from another. Besides, some other studies also reported shock reflection along with their interactions for different geometric configurations[10-18].

Skews[19] in 1967 studied the behavior of flow within the perturbed region caused by shock diffraction by varying Mach number $M_i$ (between 1.0 to 5.0) and convex corner angle $\theta$ (between 15° to 165°). He showed that parameters such as slipstream position, the tail of expansion fan, velocities of the contact surface, and secondary shock became independent of corner angles for angles greater than 75°.

Sun and Takayama[20] numerically measured the vorticity production due to the shock-wave diffraction near sharp convex corners for similar parametric space as that of Skews[19]. They reported that the rate of vorticity production sharply increases as the corner angle is varied from 15° to 45° and hardly increases for corner angles over 90°. They also observed that for a given wall angle, the shock strength encourages vorticity production. They argued that a large portion of total vorticity is represented by the slipstream than the baroclinic effects and proposed an analytical model for its estimation. Tseng and Yang[21] numerically investigated vorticity production in shock-wave diffraction around a convex corner and subsequent interaction between reflected shock-wave and the main vortex. They found the influence of reflection on the rate of vorticity production and that the degree of the production depended on the strength of the incident shockwave and the diffracting angle.

Quinn and Kontis[22] investigated shock-wave diffraction at $M_i = 1.46$ around a 172° corner angle both numerically and experimentally. They captured KH-instabilities for a very fine mesh grid. Gnani et al.[23] performed a qualitative study on shock-diffraction around sharp and curved splitter. Chaudhuri and Jacob[24] presented a detailed transient flow analysis for shock-wave diffraction over a sharp splitter plate using probability density functions of various enstrophy equation parameters and invariants of the velocity gradient tensor.

Schardin's famous flow visualization experiment[25] showed that moving shock ($M_i = 1.3$) interaction with simplest of the geometry (finite triangular wedge) can be used to study a highly complex flow field which involves several shock wave phenomena such as reflection, diffraction, shock-vortex interaction. Numerically Sivier et al.[26] and Halder et al.[27] explored the shock vortex interactions involved in Schardin's problem.

The brief literature survey involving shock-wedge interaction studies presented above suggests that focus of most of the studies were solely devoted towards understanding the shock wave phenomena in a reductionist manner, i.e. studying an only particular aspect of the flow phenomena such as either reflection[4, 28-31] or diffraction phenomena[16,19,32-34] or shock vortex interactions[35-39] in an isolated manner. Inspired by Schardin's experiment[25] discussed above which provides an opportunity to holistically study the shockwave phenomena, we present our study which tries to address two questions: (i) How do the incident shock Mach number and wall angle (two most important parameter in shock- infinite wedge interaction) affect the flow physics in Schardin's shock-finite wedge interaction setup, and (ii) If the moving shock drives the finite wedge forward what happens to its shock-wave configurations? Especially in light of the influence of incident shock Mach number and wall angle.

Three different incident shock Mach numbers ($M_i = 1.3, 1.9, 2.5$) and second wall angles ($\theta_{2w} = 60°, 90°, 120°$ as shown in **Fig.1**) are considered for the study while keeping the angle of the first wall fixed. The parametric study is performed for both stationary conditions as well as shock driven conditions, thereby presenting a total of 18 cases to shed light on the effect of the parameters $M_i, \theta_{2w}$ on the resulting flow physics.

With the advent of computational tools, especially with the development of frameworks like level set methods[40-42] and immersed boundary methods[43-49] in the past two decades, it is possible today to model highly complex unsteady shock interactions with arbitrarily complex and moving/deforming geometries[50-52]. We use our recently developed sharp interface based immersed boundary framework[53-57] for numerically studying these shock-finite wedge interaction phenomena.

The paper is organized as follows. Section II presents the details of the solver and computational approach used. Section III presents the results and discussions and conclusion in section IV.

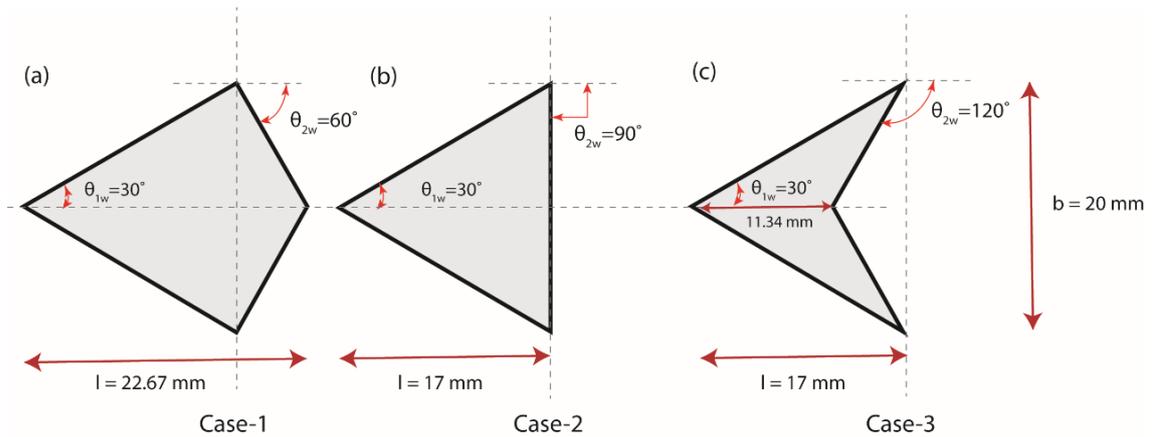

**Figure 1:** Schematic of three triangular wedge configurations with varying wall angles (a) Case-1: $\theta_{2w} = 60°$ (b) Case-2: $\theta_{2w} = 90°$ (c) Case-3: $\theta_{2w} = 120°$

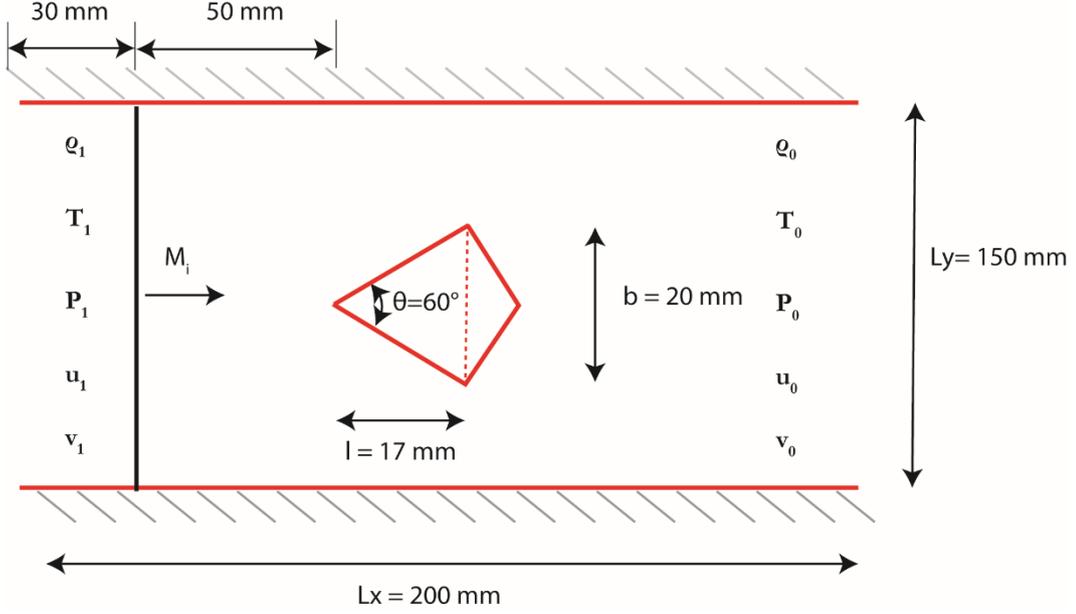

Figure 2: Schematic of Shock-Wedge Interaction.

## II. Numerical Methods

### A. Governing Equations

We use our in-house[53-55, 58-64] finite volume based multi-block compressible flow solver. It solves for 3D Favre averaged preconditioned N-S equations given below

$$\Gamma \frac{\partial}{\partial \tau} U + \frac{\partial}{\partial t} Q + \frac{\partial}{\partial x}(E - E_v) + \frac{\partial}{\partial y}(F - F_v) + \frac{\partial}{\partial z}(G - G_v) = S \quad (1)$$

Where $\Gamma$ is the preconditioning matrix given by Weiss and Smith[65] and U is the vector of primitive variables

$$U = [p, u, v, w, T]^T \quad (2)$$

Q is the conserved variable vectors. E, F, G are inviscid fluxes while $E_v$, $F_v$, $G_v$ are viscous fluxes. S is the source term.

### B. Solver Details

The solver uses the Low Diffusion Flux Splitting scheme[66] for discretizing convective fluxes. A fifth-order weighted essentially non-oscillatory (WENO) scheme[67] is utilized for inviscid flux reconstruction. The scheme is used for their capability of achieving higher spatial accuracy and better shock resolutions, especially where strong discontinuities are present. A central difference scheme is used for viscous fluxes. Time marching is through a dual time-stepping approach. A second order backward three-point differencing is used for discretizing physical time step while explicit Euler is used for local pseudo time stepping. Parallel processors communicate using MPI.

For handling flows involving complex geometries and complex fluid-structure flow phenomena, the solver utilizes a sharp interface immersed boundary framework[53-57]. The framework uses a hybrid of HCIB approach, and ghost cell approach for flow reconstruction, which ensures mass conservation and suppression of spurious oscillations when moving body problems are involved. Sharp interface methods usually face serious limitations when encountering geometries involving sharp edges. Sharp edges have infinite curvature where the surface normal is not well defined. This adversely affects both the node classification algorithm and flow reconstruction near sharp edges. The present solver uses a novel reconstruction algorithm[56, 57], that utilizes ray casting algorithm[68] and a reconstruction strategy that defines angle-weighted pseudo-normals[69] for vertex and edges where normals are not defined. This ensures the immersed body is well oriented at all the points on the surface. Based on whether a given node (which is to be reconstructed) falls near the vertex, edge, or face, on the solution reconstruction algorithm chooses the respective local normal as the direction of reconstruction.

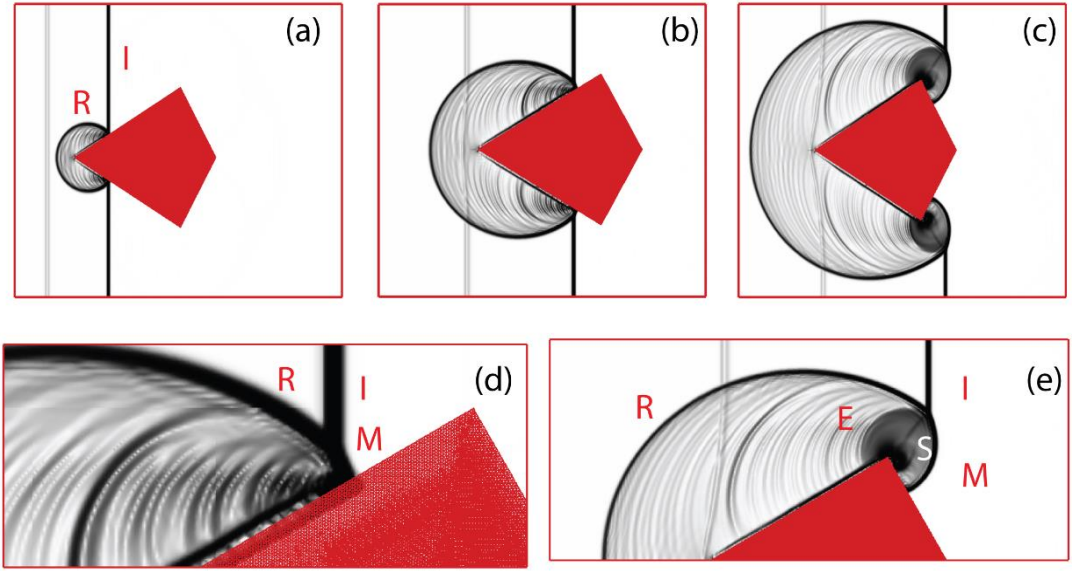

**Figure 3:** Shock Reflection and Diffraction for case 1: $\theta_{2w} = 60°$ at $M_i$ = 1.3. (a) t = 47.25μs (b) t = 70.875μs  (c) t = 94.5μs (d) Enlarged view of three shock system at  t = 70.875μs  (e) Enlarged view of shock diffraction at  t = 94.5μs .

## III. Results and Discussion

### A. Moving Shock Interactions with Stationary Wedge

#### 1. Computational Domain Details

The schematic of the computational domain is presented in **Fig.2**. The computational domain is of size $200mm \times 150mm$. The dimensions are adopted from the work of A.Chaudhuri et al[43]. The domain is meshed with computational nodes of $1170 \times 850$. The mesh in the immersed body region is uniformly refined with a mesh resolution of $\Delta x = l/200$. The moving shock is placed initially at $x = 30\ mm$ from the origin and the wedge at $x = 80mm$. The post shock flow is initialized with following conditions.

$$\rho_0 = 1.1825 kgm^{-3}, \gamma = 1.4, u_0 = 0.0, v = 0.0, P_0 = 0.05 MPa, T_0 = 300K$$

The pre-shock flow field, which is a function of post-shock conditions and moving shock strength, is determined using standard Rankine-Hugoniot relations[70]. The inlet condition is set identical to the pre-shock condition while at outlet convective boundary condition is used. The top and bottom domains are set to free stream conditions.

#### 2. Shock Wave Reflection and Diffraction

Shock wave interaction with a triangular wedge can be divided into two processes: (a) Interaction of incident shock (I) with the first planar surface $\theta_{1w}$ results in a reflected shock wave (R), which propagates upstream cylindrically, forming a regular reflection (see **Fig.3**). As it reaches the end of the surface, the two-shock system evolves into three shock due to the formation of Mach stem (M); (b) As the shock wave moves past the wedge, it encounters a sharp convex edge leading to shock diffraction. This generates expansion fan (E) with diffracted shock M. The slip layer too coils at the sharp corner of the wedge resulting in a vortex. The slip line S now emanates from the triple point.

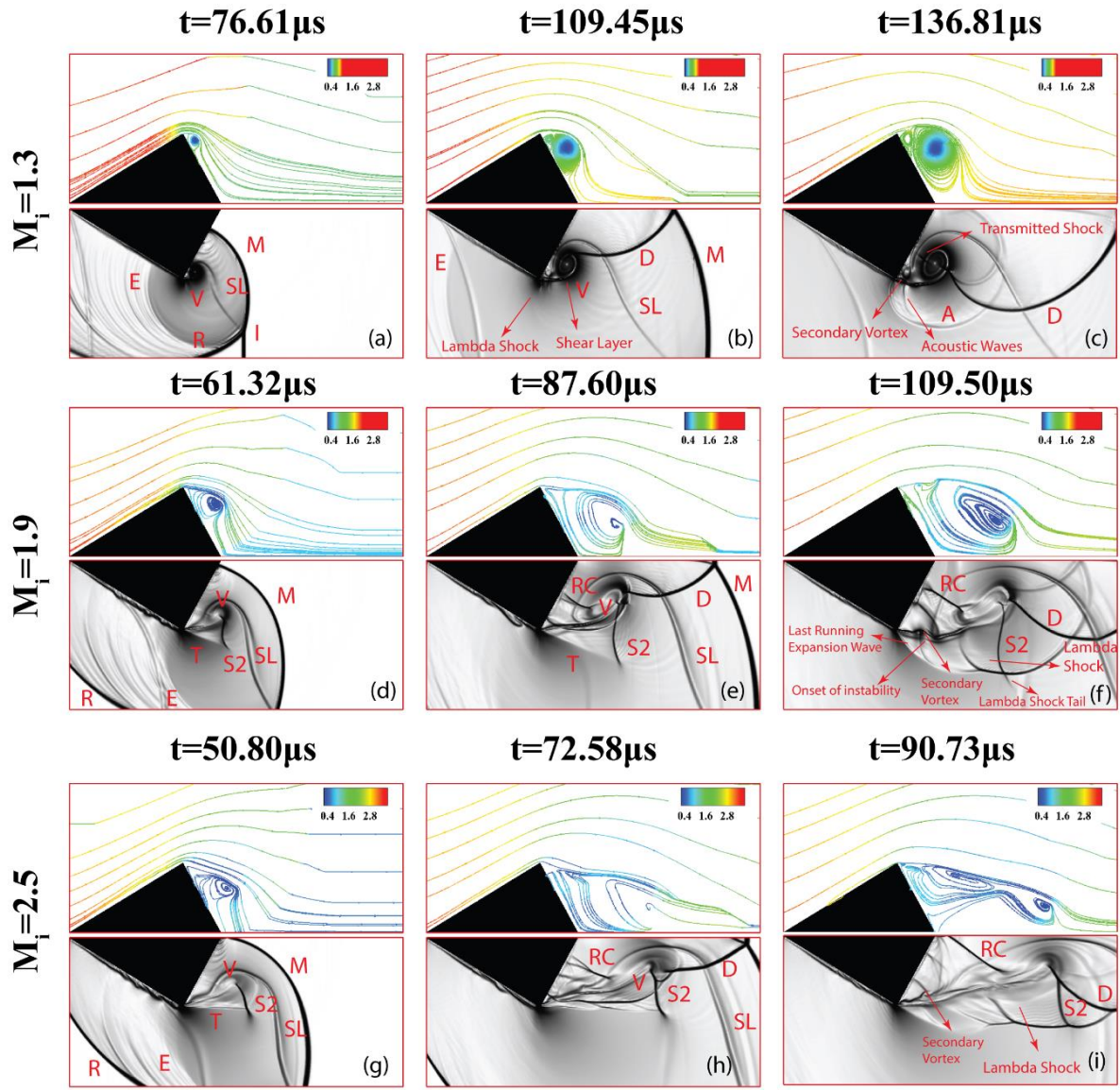

**Figure 4**: Comparative plot showing the evolution of vortex generated due to shock diffraction at different incident Mach number $M_i$ for case 1: $\theta_{2w} = 60°$. First Row: (a-c) $M_i = 1.3$; Second Row: (d-f) $M_i = 1.9$; Third Row: (g-i) $M_i = 2.5$  R:Reflected Shock; I:Incident Shock; M:Mach Stem; E:Expansion Fan; SL: Slip Layer; V:Vortex; D: Decelerated Shock; A: Accelerated Shock; S2: Secondary Shock; RC: Recompression wave; T: Terminator; TS: Transmitted Shock

The reflected shock on interacting with vortex is scattered into accelerated and decelerated shock waves. Further, the decelerated shock wave interacts with the vortexlets developed along with the slip layer (in case of inviscid flows) or shear layer (in case of viscous flows). This leads to a further generation of acoustic waves and attenuation of decelerated shock. While the first process, in general, is observed in all the simulated cases (as $\theta_{1w}$ is the same), there is a considerable variation in the second process with the variation of incident shock Mach number and wall angle.

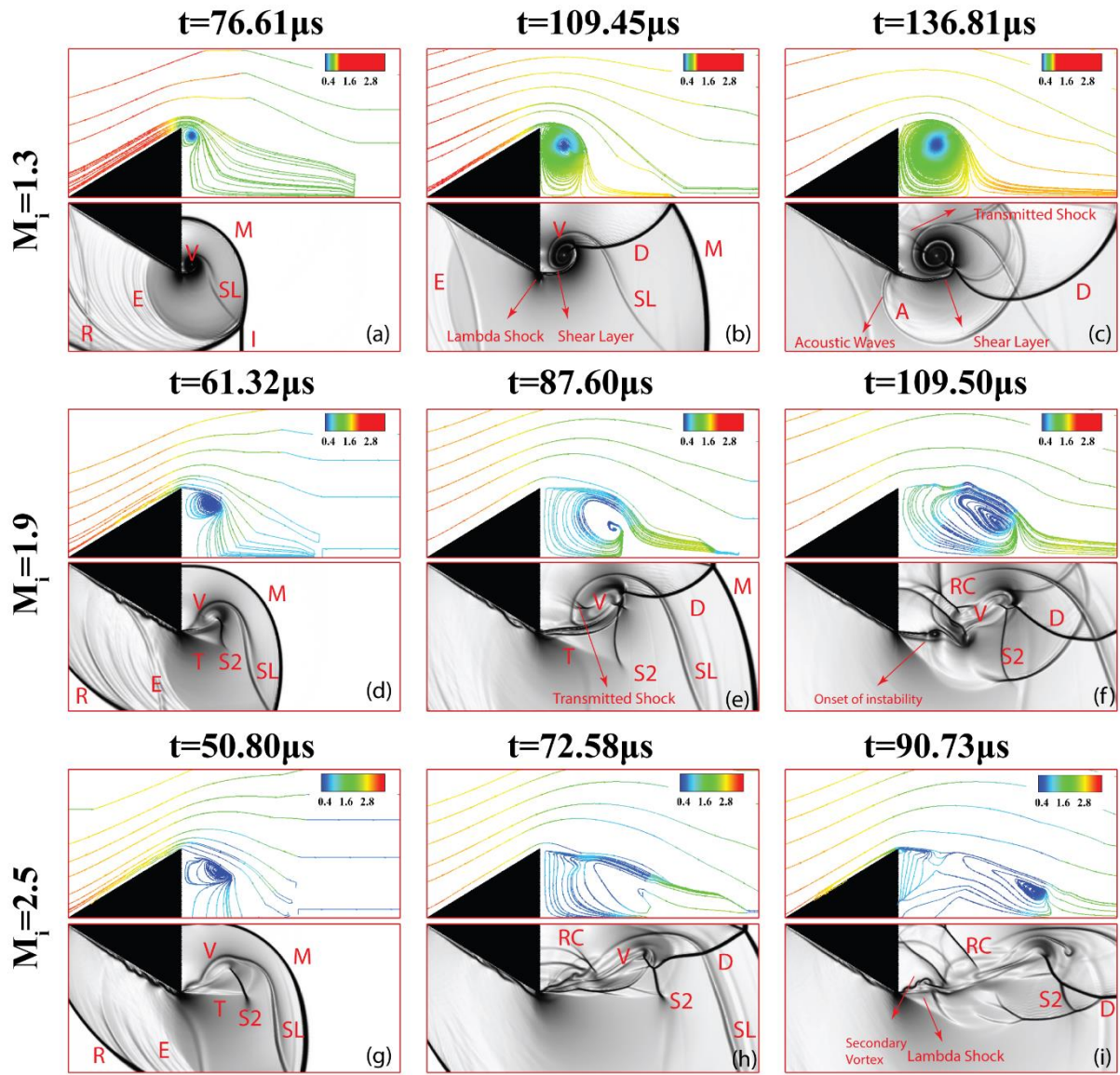

**Figure 5**: Comparative plot showing the evolution of vortex generated due to shock diffraction at different incident Mach number $M_i$ for case 2: $\theta_{2w} = 90°$. First Row: (a-c) $M_i = 1.3$; Second Row: (d-f) $M_i = 1.9$; Third Row: (g-i) $M_i = 2.5$; R:Reflected Shock; I:Incident Shock; M:Mach Stem; E:Expansion Fan; SL: Slip Layer; V:Vortex; D: Decelerated Shock; A: Accelerated Shock; S2: Secondary Shock; RC: Recompression wave; T: Terminator; TS: Transmitted Shock

**Figs.4-6** depicts the comparative plot showing the evolution of vortex generated due to the shock diffracted at the convex corner of different cases $\theta_{2w} = 60°, 90°, 120°$ at different incident Mach numbers ($M_i = 1.3, 1.9, 2.5$). While the upper part of the image shows streamline patterns, the lower part of the image shows a numerical schlieren image showing the shock patterns involved. Consider **Fig.4a-c** corresponding to $\theta_{2w} = 60°$ at $M_i = 1.3$ where the weak shock on diffraction at wedge corner produces vortex whose Mach number is subsonic. **Fig.4a** shows reflected shock (R) expanding towards the front and Mach stem (M) diffracted around the tip of the wedge. The slip layers (SL) coil around to form vortices (V). An expansion fan (E) is also visible. The incident shock (I) is perpendicular to the direction of the motion of the shock. As the incident shock (I) moves

forward, the reflected Mach stem (RM) from the upper surface interacts with the vortex (V) at the lower tip of the wedge, as shown in **Fig.4b.** As reflected Mach stem (RM) impinges on the vortex (V), it gets scattered into accelerated shock (A) and decelerated shock (D). **Fig.4c** illustrates that decelerated shock (D) penetrates the vortex core to develop a transmitted shock (TS) while one end of accelerated shock (A) is seen to be reflected from the vertical wedge wall. Noticeably, the presence of secondary vortex near the convex corner is a characteristic of viscous flow.

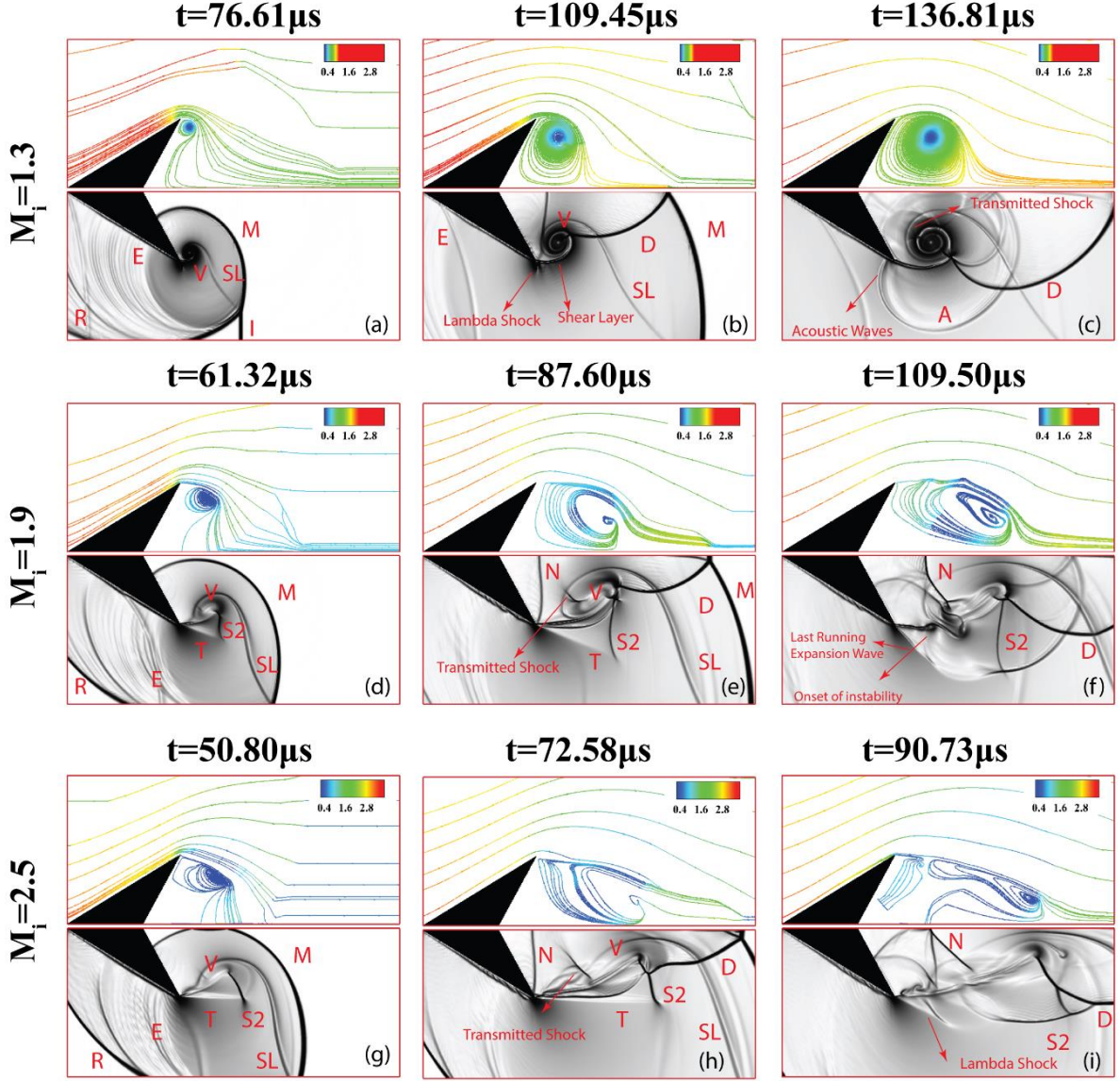

**Figure 6**: Comparative plot showing the evolution of vortex generated due to shock diffraction at different incident Mach number $M_i$ for case 3: $\theta_{2w} = 120°$. First Row: (a-c) $M_i = 1.3$; Second Row: (d-f) $M_i = 1.9$; Third Row: (g-i) $M_i = 2.5$; R:Reflected Shock; I:Incident Shock; M:Mach Stem; E:Expansion Fan; SL: Slip Layer; V:Vortex; D: Decelerated Shock; A: Accelerated Shock; S2: Secondary Shock; RC: Recompression wave; T: Terminator; TS: Transmitted Shock

At $M_i = 1.9$ (**Figs. 4d-f**), the diffracted region is completely supersonic. Presence of centered expansion fan with a well-pronounced tail called terminator (T) and a secondary shock (S2) that is bounded by the terminator (T) and slipstream (SL), as shown in **Fig.4d,** is characteristic of the flow field at higher incident Mach number. The evolution of the flow field suggests the formation of recompression waves (RC) that connect the cortex core to the wall and the shear layer undergoing KH instability (**Fig.4e-f**). With a further increase in the incident Mach number to $M_i = 2.5$, the vortex becomes further elongated and difficult to identify its core. Corresponding flow field evolution (**Figs.4g-i**) again confirms the formation of the secondary vortex at the convex corner of the wedge and the shear layer joining the main vortex undergoing KH instability. A similar diffraction process is observed for $\theta_{2w} = 90°$ (**Fig.5**) and $\theta_{2w} = 120°$ (**Fig.6**). It can be noticed that both the cases do not form any secondary vortices at $M_i = 1.3$ and $1.9$ near the convex corner.

### 3. Vortex Dynamics

In order to investigate the vortex dynamics, total circulation $\Gamma$ is calculated from the expression $\Gamma = \oint u dl$ and is plotted against time in **Fig.7a-c**. As seen in the figure, with increasing Mach number the strength of the circulation increases in magnitude irrespective of the value of $\theta_{2w}$. This is consistent with the observation of Sun and Takayama[16]. Also, notice that circulation increases with the time almost linearly as expected and is slightly bent. This has also been observed by Tseng and Yang[17]. For $\theta_{2w} = 90°, 120°$, the trend of the circulation and magnitude of the circulation for a given Mach number $M_i$ almost remains the same. A slight offset is seen between them for $M_i = 1.3$. For $\theta_{2w} = 60°$, the presence of wall very near to vortex as shown in **Fig.4** inhibits the growth of the vortex freely unlike the case of two other wall angles $\theta_{2w}$. Thus, the wall vortex interaction as well as shock vortex interaction results in loss of circulation as the vortex evolves. This is very much apparent in the case of $M_i = 1.3$. For higher Mach number the circulation drops and then recovers after a while. This can be attributed to the presence of strong shocks and its interaction with the shear layer and vortex present in the flow, leading to flow instabilities, which enhances the vorticity. Evidence for vorticity generation will be presented a little later while discussing vorticity transport equation budget terms.

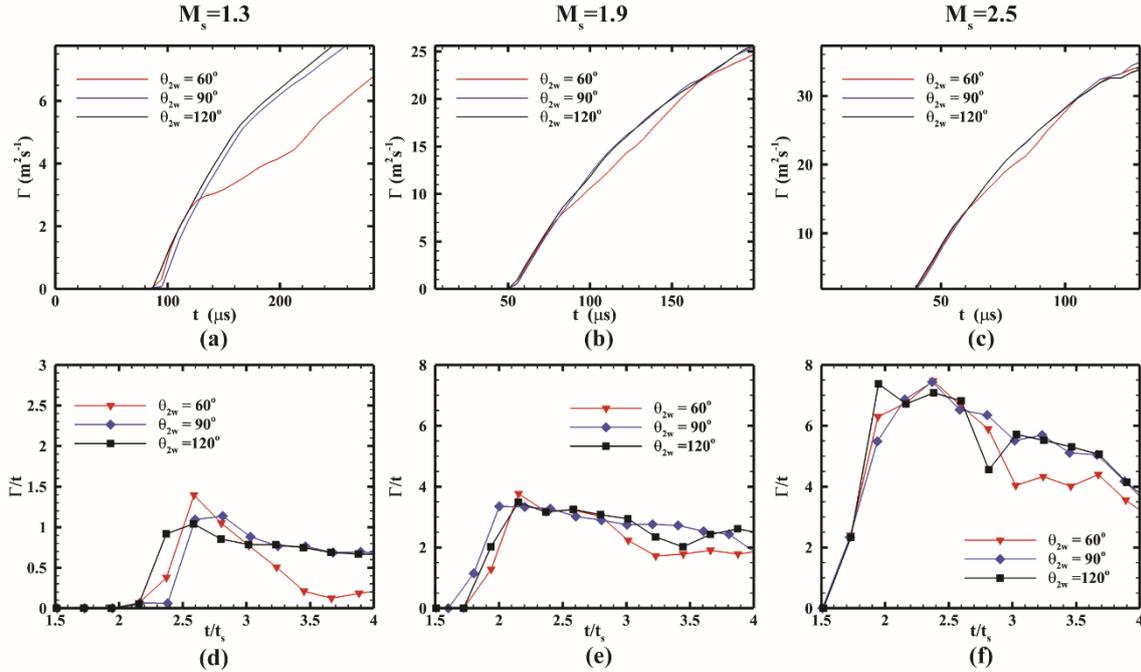

**Figure 7:** First row (a-c) shows circulation $\Gamma$ vs. time. Second row (d-f) shows the normalized rate of circulation production $\Gamma/t$ vs. non-dimensionalized time

**Figs.7d-f** present the normalized rate of circulation production $\Gamma/t$ with respect to non-dimensional time ($t/t_s$). Here, $t_s$ is the time taken by the shock to move through the characteristic length of the wedge, and the rate of circulation production ($\Gamma/t$) is normalized by $RT_0$, where R is the gas constant for air and $T_0$ is the initial post-shock temperature (300K)[16]. The rate of circulation production is a function of incident shock $M_i$. For a given $\theta_{2w}$, the rate of circulation production increases as the $M_i$ increases, as seen from **Fig.7d-f**. Similarly, trends in the rate of circulation production are similar for $\theta_{2w} = 90°, 120°$. For $\theta_{2w} = 60°$ as expected, the rate of circulation production term is lesser compared to the other two cases. The wall angle acts as a hindrance to the vorticity production.

Understanding the mechanisms involved in the vortex dynamics can be developed by closely examining the terms involved in the vorticity transport equation (VTE). The equation can be written as

$$\underbrace{\frac{D\omega}{Dt} = (\omega.\nabla)u}_{VSG} - \underbrace{\omega(\nabla.u)}_{VSC} + \underbrace{\frac{1}{\rho^2}(\nabla\rho \times \nabla P)}_{BAR} + \underbrace{\nabla \times \left(\frac{\nabla.\tau}{\rho}\right)}_{DFV} \qquad (3)$$

where the first term VSG in R.H.S represents the vorticity stretching due to flow velocity gradients (VSG=0 for 2D flows), the second term VSC is vorticity stretching due to flow compressibility. The third term, BAR is the baroclinic term that accounts for the changes in the vorticity due to the presence of pressure and density gradients. The final term DFV represents the vorticity diffusion due to the viscous effects.

**Fig.8** reports the temporal evolution of volumetric average of VTE budgets for $\theta_{2w} = 60°, 90°, 120°$ at $M_i = 1.9$. It can be observed from the figure that diffusion effects dominate the vorticity evolution followed by vorticity stretching phenomena. The baroclinic term contributes very little on average sense. It is observed that the temporal evolution of VTE follows a similar trend for $\theta_{2w} = 60°, 90°$ except towards the end where the difference becomes prominent. VSC term's (**Fig.8a**) evolution shows the presence of a valley and peak at the initial time (till $t/t_s$= 2.5) and is relatively flat for the rest of the time (till $t/t_s$= 5.5) before it again peaks towards the end. BAR term's (**Fig.8b**) evolution shows that for initial times till $t/t_s$= 2.0, they are relatively flat, and for the rest of the time, it shows random fluctuations. DFV term's (**Fig.8c**) shows that it is relatively flat for half of the time (till $t/t_s$= 3.0) shows fluctuations for the rest. The total contribution of these terms to VTE is presented in **Fig.8d.** The temporal evolution is almost constant and of the same magnitude till the time instant $t/t_s$ = 3.2, after which the fluctuations in quantity is observed. The later time corresponds to time instances that involve complex shock-vortex interactions. A temporal average of the magnitude of total VTE term will show that vorticity production on average was least in for $\theta_{2w} = 120°$ when compared to $\theta_{2w} = 60°, 90°$

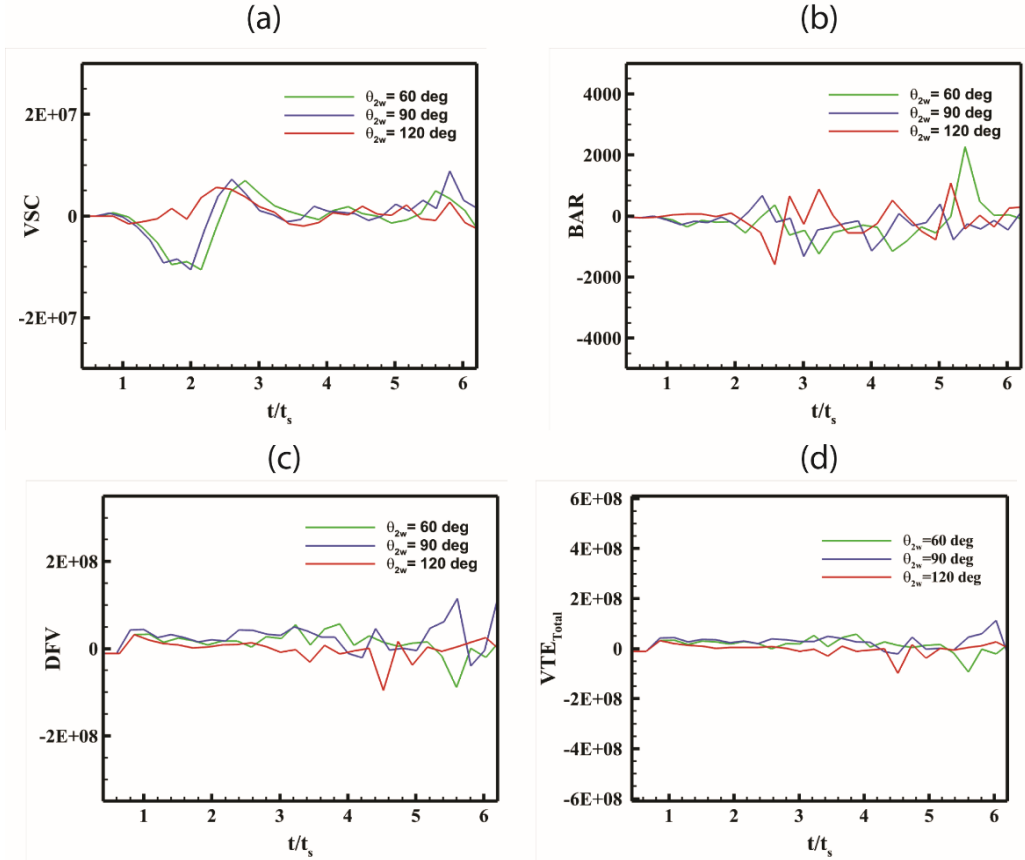

**Figure 8**: Vorticity Transport Budget temporal evolution for $\theta_{2w} = 60°, 90°, 120°$ at $M_i$=1.9. (a) VSC (b) BAR (c) DFV (d) VTE$_{Total}$

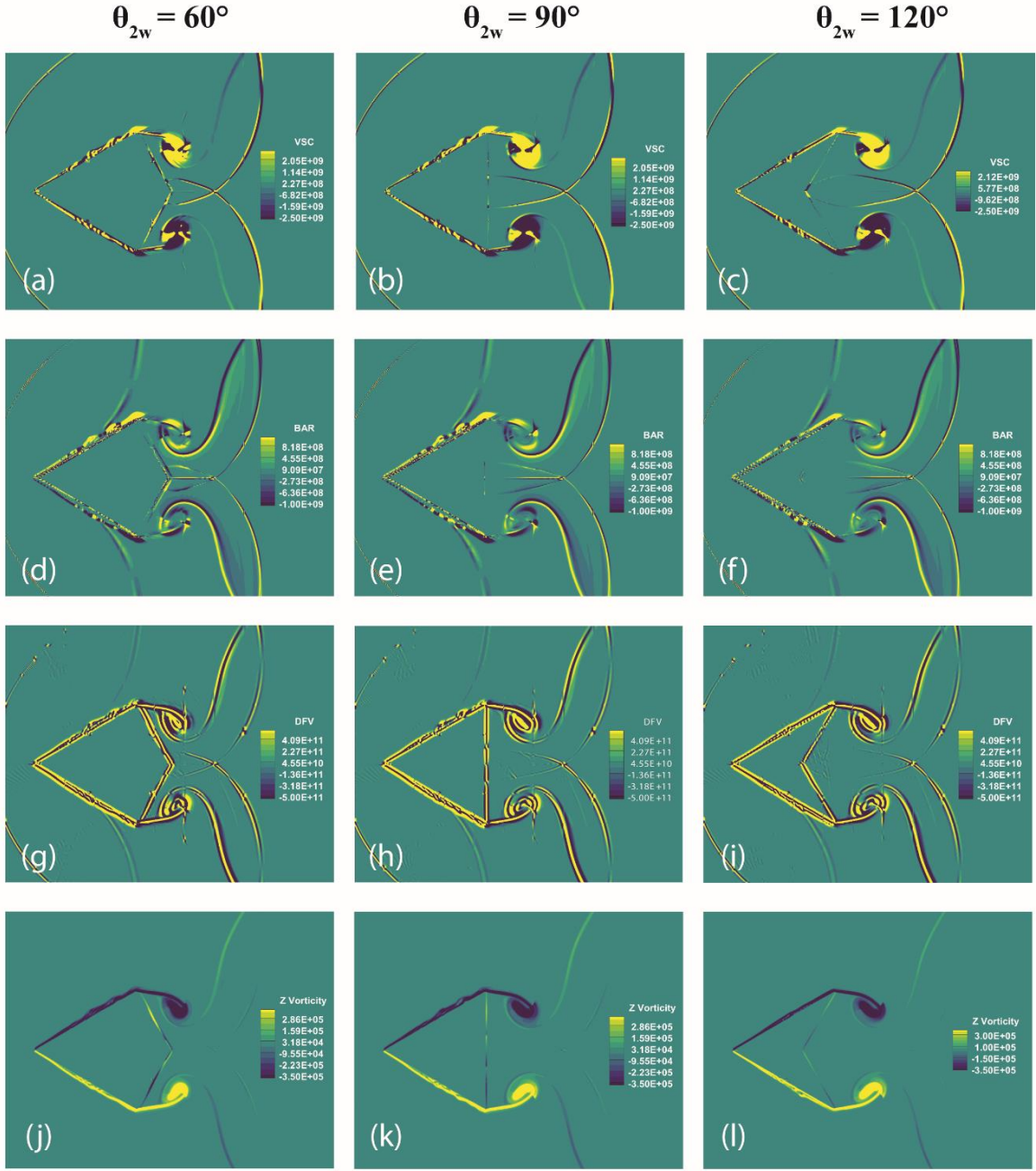

**Figure 9**: Vorticity Transport Budget for $\theta_{2w} = 60°, 90°, 120°$ at $M_i$=1.9. First row: (a-c) VSC; Second row: (d-f) BAR; Third row: (g-i) DFV; Fourth row: (j-l) Z-vorticity; All the images correspond to time instance t=129µs

**Fig.9** shows VTE budget contours for all the three cases $\theta_{2w} = 60°, 90°, 120°$ corresponding to time instance $t/t_s$=3.2, which is before the shock vortex interaction happens. The VSC contour plot (**Fig.9a, 9b, 9c**) shows the presence of locally stretched structures around the vortex core due to compressibility effects. The BAR contour plot (**Fig.9d, 9e, 9f**) captures the misalignment of pressure and density gradients, which in turn shows the vorticity generated due to the presence of contact discontinuity, slip lines, and reflected shock structures. The DFV contour plot (**Fig.9g, 9h, 9i**) depicts the diffusion of vorticity due to the viscous effects. As pointed out in **Fig.8**, the magnitudes of DFV dominates, and BAR contribution is the lowest.

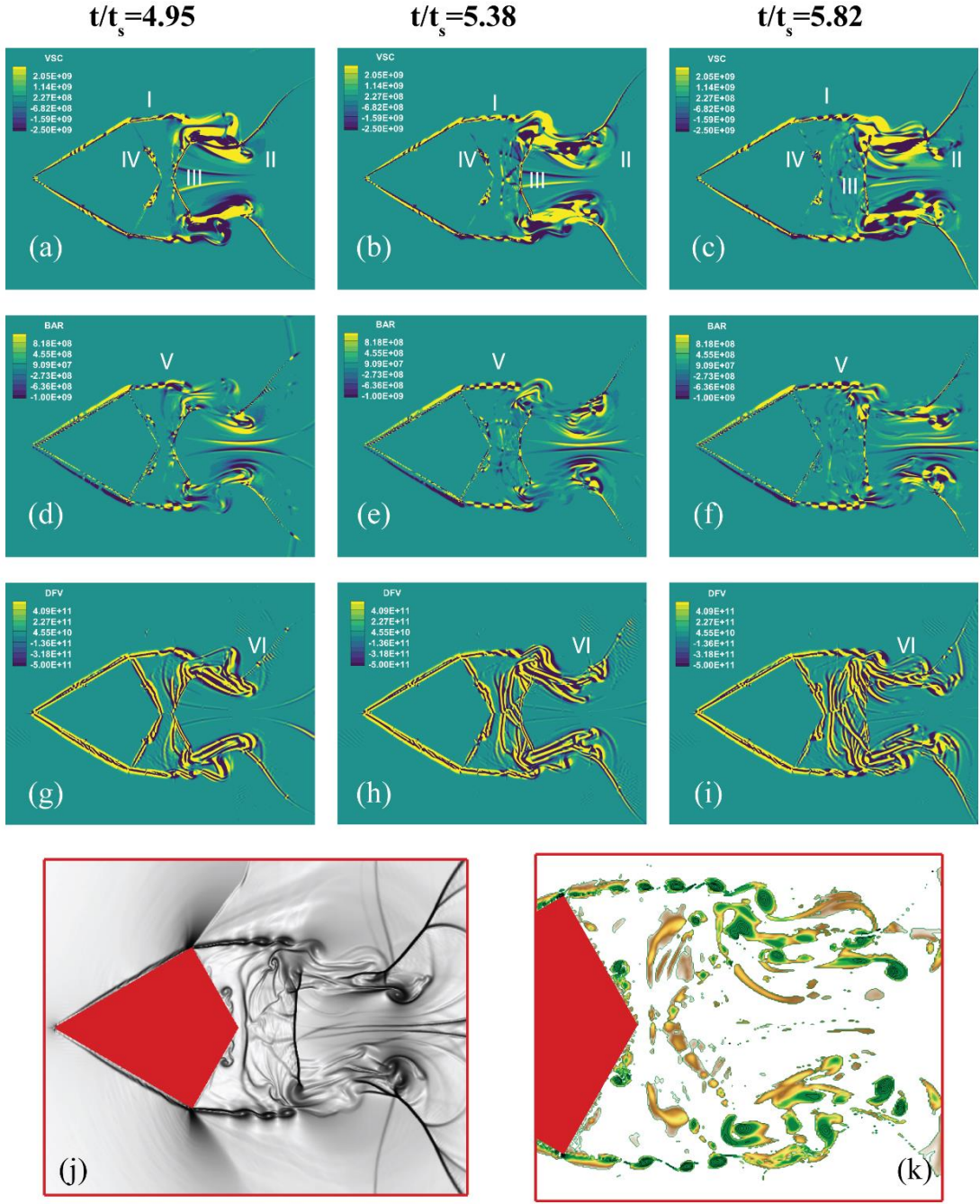

**Figure 10**: Vorticity Transport Budget for $\theta_{2w} = 60°$ at $M_i = 1.9$; First row: (a-c) Vorticity stretching due to compressibility (VSC); Second Row: (d-f) Baroclinic term (BAR); Third Row: (g-i) Vorticity Diffusion term due to viscous effects (DFV); Last Row: (j) Numerical Schlieren image showing flow structures at t=284.85μs. (k) Vortex structures obtained from $\lambda_{ci}$-criteria

The evolution of shock-vortex structures for two cases $\theta_{2w} = 60°, 120°$ corresponding to $M_i = 1.9$ are presented in **Fig.10-11** at three-time instances (t=237.35μs, 261.12μs, 284.85μs). These instances, which are later than the instant shown in **Fig.9**, depicts the shock-vortex interaction and its evolution. The VSC contour plots (**Fig.10a, 10b, 10c**) show the vorticity stretching present in shear layer (I), around the vortex core (II), near the embedded shock (III) and around the wall vortex (IV). The presence of KH instability can be inferred from alternating vorticity stretch patches in the shear layer (I). The initially curved embedded shock (III) evolves into a normal embedded shock as it moves forward. Interaction between the shock and vortex leads to the distortion and breakdown of a large vortex (II) into small vortices are also observed. The BAR contour plots (**Fig.10d, 10e,**

**10f**) illustrate the vorticity generated due to the unequal acceleration resulted from the misalignment of pressure and density gradients. This essentially will cause lighter density fluid to accelerate faster than higher density fluid resulting in the shear layer (V). The DFV contour plots (**Fig.10g, 10h, 10i**) depicts the spread of vorticity (VI) due to actions of viscosity.

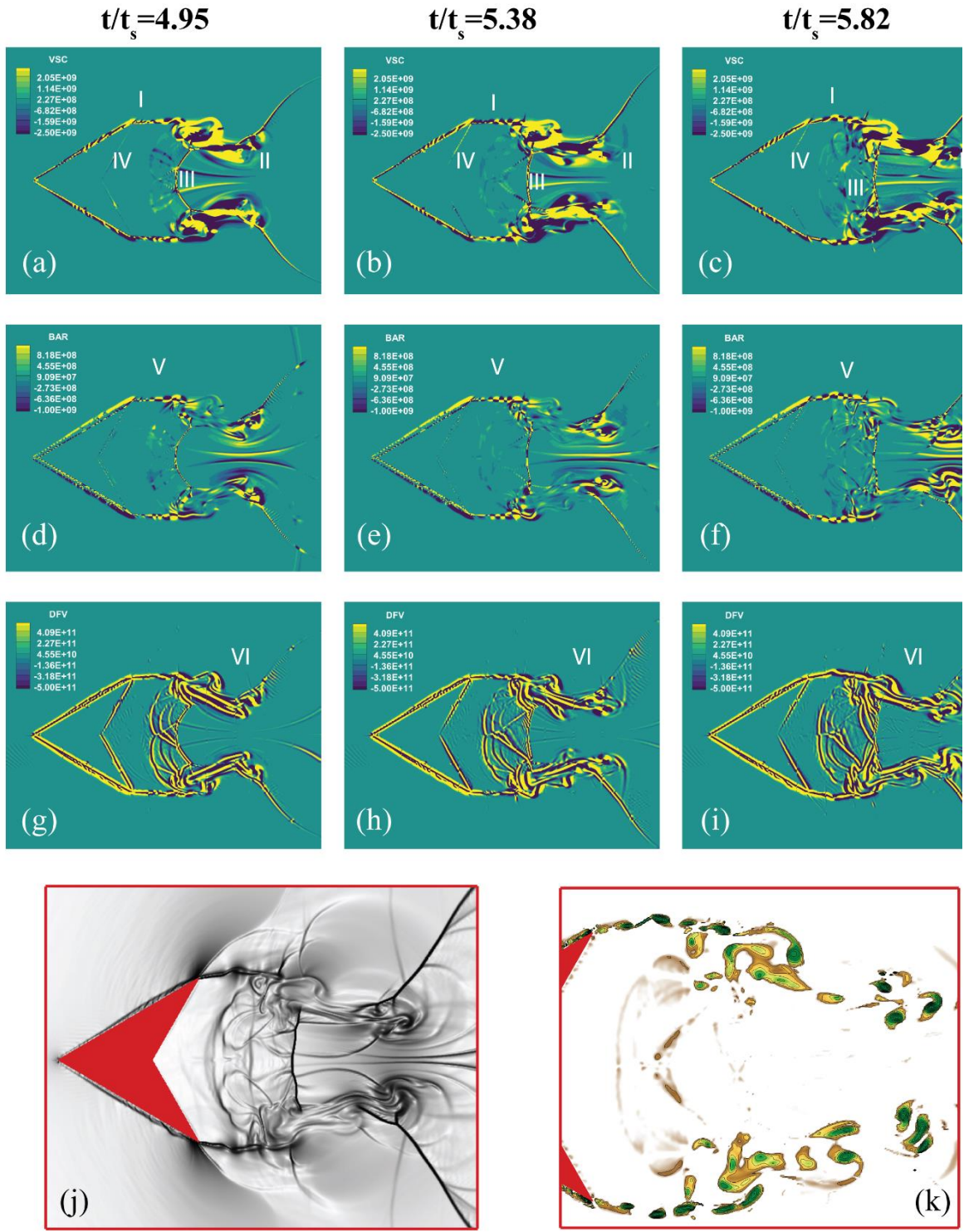

**Figure 11**: Vorticity Transport Budget for $\theta_{2w} = 120°$ at $M_i = 1.9$; First row: (a-c) Vorticity stretching due to compressibility (VSC); Second Row: (d-f) Baroclinic term (BAR); Third Row: (g-i) Vorticity Diffusion term due to viscous effects (DFV); Last Row: (j) Numerical Schlieren image showing flow structures at t=284.85μs. (k) Vortex structures obtained from $\lambda_{ci}$-criteria

The numerical schlieren image of the case $\theta_{2w} = 60°$ is shown in **Fig.10j** for time instant t=284.85μs. The image depicts a decay of the main vortex into multiple small vortices due to the action of multiple embedded and reflected shocks. The shear layer too is broken into multiple small vortices. **Fig.10k** depicts the visualization of vortices present in the wake with the help of swirl strength $\lambda_{ci}$-criteria. One can notice the presence of a vortex structure at the wedge wall. This is due to the shock-boundary layer interaction. **Fig. 11** illustrates similar images corresponding to $\theta_{2w} = 120°$. In contrast to the $\theta_{2w} = 60°$ case, **Fig.11k** shows no wall vortex interaction. The spread of the vortices two are away from the wedge centerline unlike $\theta_{2w} = 60°$(**Fig.10k**). This once again demonstrates that vorticity generation is minimum for $\theta_{2w} = 120°$ when compared to other cases (as we saw from **Fig.8** VTE plots).

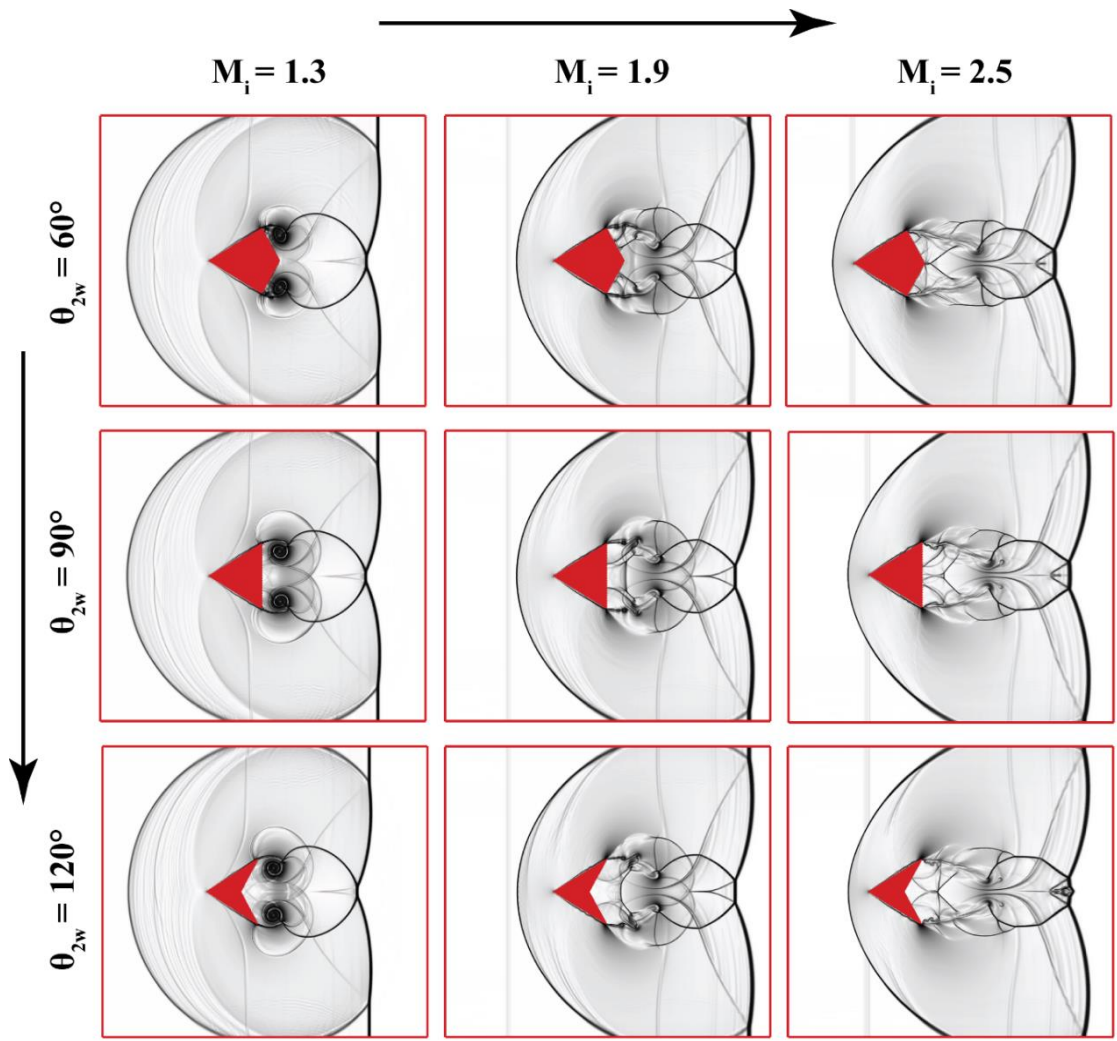

**Figure 12:** Array of numerical Schlieren images showing flow structures resulting from complex interactions of moving shock-stationary wedge cases. Left to right shows the variation of flow structure with increasing incident shock Mach number while top to bottom shows the flow structure variation with an increase in corner angle. The three columns $M_i$ =1.3, 1.9, 2.5 correspond to the physical times t = 283.5μs, 199.96μs, 152μs respectively

Flow structure developed through such complex shock/vortex interaction for all the nine possible cases is shown in **Fig.12**. Different features are observed for different $\theta_{2w}$ and $M_i$. The flow structures are symmetric between upper and lower halves of the wedge. As the Mach number increases, the area between the front tip of the wedge and bow shock shrinks, which shows that bow shock (sonic wave) covers a smaller distance relative to the incident shock wave moving downstream. A closer look at the wake structure shows that corner walls, vortex, and shock structures interactions result in the different flow structures in the immediate wake, especially with respect to the embedded shock structures present in the vortices.

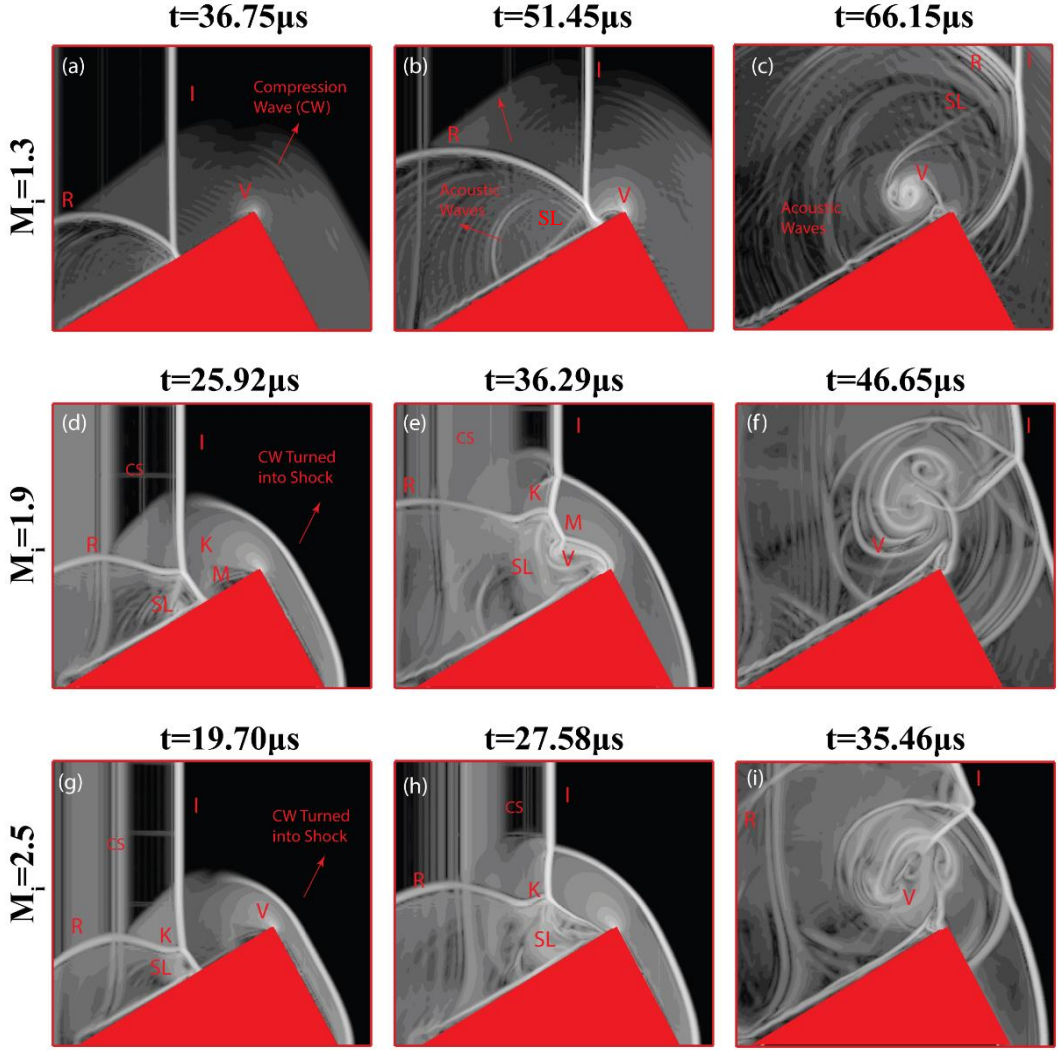

**Figure 13**: Numerical Schlieren image of a case $\theta_{2w} = 60°$ corresponding to three-time instances t=19.70µs, 27.58 µs, 46.65 µs at three different incident Mach numbers. First Row: (a-c) $M_i = 1.3$; Second Row: (d-f) $M_i = 1.9$; Third Row: (g-i) $M_i = 2.5$;

## B. Shock Wave Interactions in Moving Shock Driven Wedge

In this section, we present the shock driven wedge cases for a similar set of parameter space ($\theta_{2w}$, $M_i$) as described in Section. IIIA ($\theta_{2w} = 60°, 90°, 120°; M_i = 1.3, 1.9, 2.5$) for the same computational domain. The initial conditions for these cases are the same as mentioned earlier. The immersed body is assumed to be rigid. The coupling between the fluid and immersed body is enabled by using the force field information generated due to the shock interaction with the wedge to update the velocity and position of the immersed body at the advanced time levels using Newtonian equations. A verlet integration approach is used for the purpose. The approach has been successfully applied in our earlier works[53-55, 64]. The present study is inspired by the works of Henshaw et al.[71] and Luo et al.[44], who performed similar studies using a cylinder. Here the primary focus is to investigate the transition of shock wave configurations as it is reflected from the inclined surface ($\theta_{1w}$) and also to study the diffraction pattern as the shock drives the wedge forward.

Consider **Fig.13** which depicts the numerical schlieren of case $\theta_{2w} = 60°$. The evolution of the flow field with time is presented for three instances (t=19.70µs, 27.58µs, 46.65µs) corresponding to three incident Mach numbers $M_i$. As the shock impinges on the wedge and moves through it, the wedge begins to accelerate from rest moving along with the shock. For $M_i$ = 1.3 case, as the wedge begins to move initially from rest, a compression wave (CW) is formed due to the motion of the wedge (**Fig.13a).** The corner of the wedge shows the generation of vorticity, which, as the flow field evolves, develops into the vortex. As the shock moves further past the wedge, the flow transitions from regular reflection to irregular reflection (Single Mach Reflection-SMR). **Fig.13b** shows the SMR with a clear presence slip line. As the shock moves towards the tip of the wedge, it interacts with the concentrated vorticity near the corner and generates acoustic waves, which then travel towards the front tip of the wedge. As the shock move past the corner (**Fig.13c),** unlike the stationary case where the shock diffraction would happen near the convex corner leading to vortex generation near the second wall ($\theta_{2w}$), the diffraction happens in the direction of the first wall $\theta_{1w}$. Also, instead of a single vortex (as in stationary case), a compressible vortex dipole is formed here, which is moving to impinge and collapse on the first wall. A lambda shaped Mach stem can also be observed in the same frame.

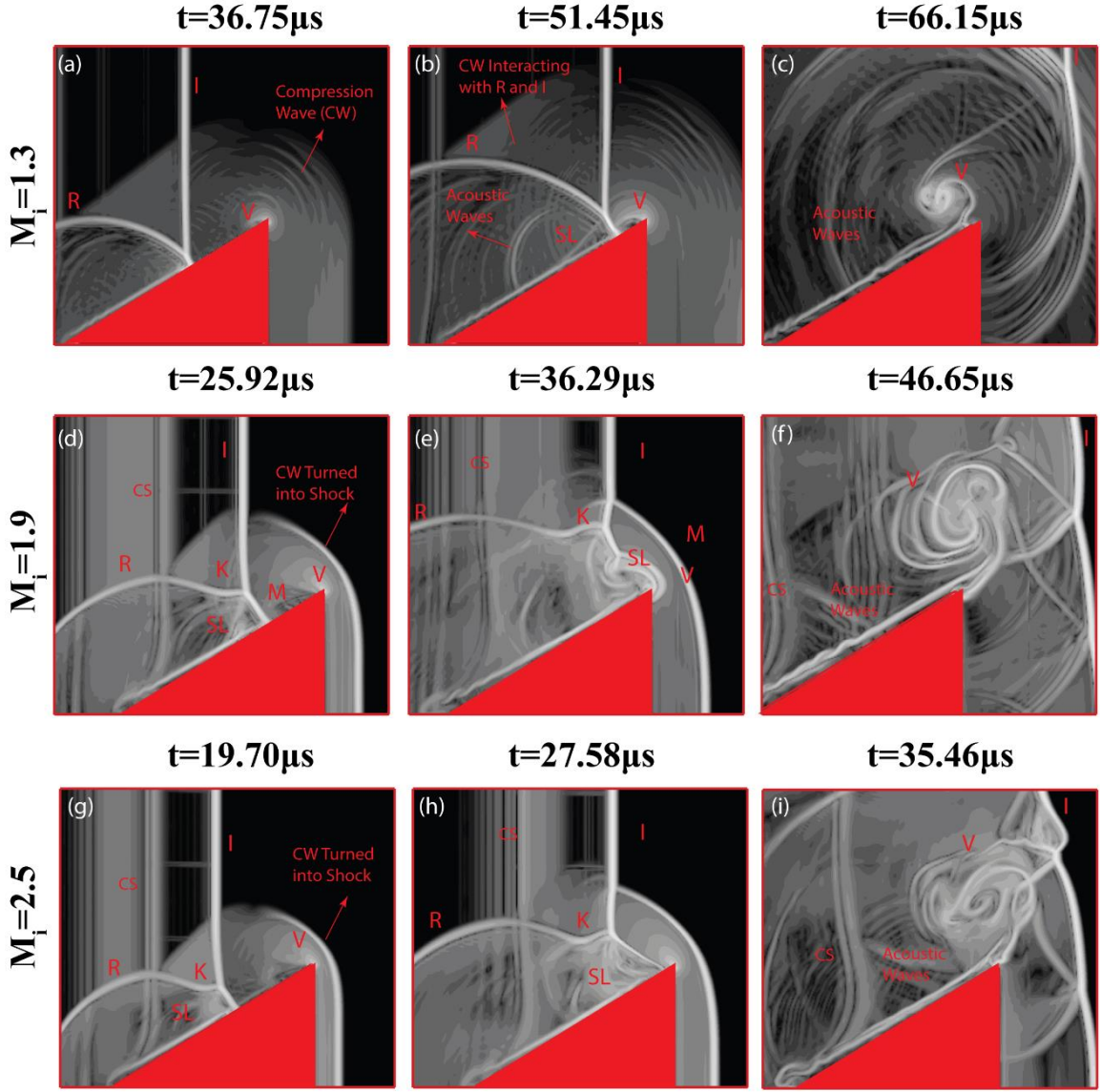

**Figure 14**: Numerical Schlieren image of a case $\theta_{2w} = 90°$ corresponding to three-time instances t=19.70µs, 27.58 µs, 46.65 µs at three different incident Mach numbers. First Row: (a-c) $M_i = 1.3$; Second Row: (d-f) $M_i = 1.9$; Third Row: (g-i) $M_i = 2.5$;

At higher Mach number $M_i$ = 1.9, **Fig.13d** suggests that the wedge too is accelerated to sonic speeds resulting in the formation of a curved shock parallel to the second wedge wall at a distance. At this incidence Mach number, the shock reflection exhibits what is termed as Transition Mach Reflection (TMR), which is an intermediate stage in the transition from Single Mach Reflection (SMR) to Double Mach Reflection (DMR). Transition Mach Reflection is characterized by the presence of Kink (K) in the reflected shock wave (R). This kink is developed as the flow field within the reflected shock wave becomes supersonic. As the shock reaches the corner, the Mach stem (M), Slip line (SL) interacts with vortex resulting in a complex flow structure (**Fig.13e**). Interaction of incident shock (I) with the shock behind the wedge, results in the development of multiple Mach stems and triple points. **Fig.13f** shows the presence of a much larger dipole vortex compared to $M_i = 1.3$. A similar flow field evolution is observed for $M_i = 2.5$ (**Figs.13g-i**). It can be noticed that the shock behind the wedge is much closer to the wedge compared to $M_i = 1.9$, and the dipole vortex formed is as strong as the ones corresponding to lower Mach numbers. Due to the strong shock interaction, the vortex dipole seems to have almost disintegrated.

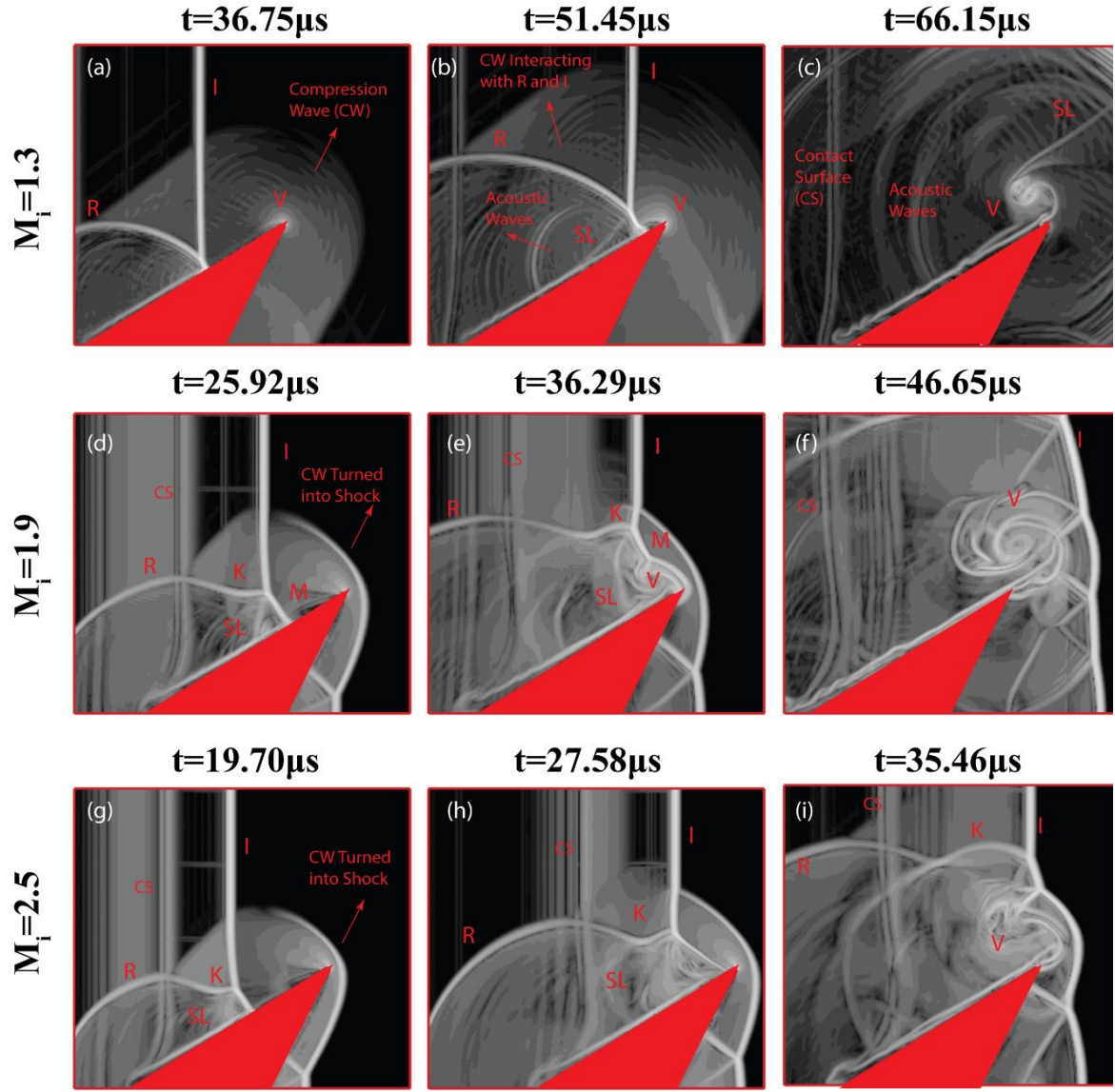

**Figure 15**: Numerical Schlieren image of a case $\theta_{2w} = 120°$ corresponding to three-time instances t=19.70µs, 27.58 µs, 46.65 µs at three different incident Mach numbers. First Row: (a-c) $M_i = 1.3$; Second Row: (d-f) $M_i = 1.9$; Third Row: (g-i) $M_i = 2.5$;

The increase of second wall angle to $\theta_{2w} = 90°, 120°$ (**Figs.14-15**), do not affect the shock reflection pattern at the first wall ($\theta_{1w}$) for different Mach numbers. They remain the same as $\theta_{2w} = 60°$. The vortex formation at the tip and shock pattern behind the wedge and their complex interaction pattern varies due to the change in the second wall angle. For $\theta_{2w} = 90°$, the second wall is perpendicular to the x-axis. The shock formed behind the wedge due to the supersonic acceleration of the wedge results in a straight shock, which only curves near the corners of the wedge (**Fig.14**). For $\theta_{2w} = 120°$, the shock behind the wedge is not a single shock but multiple shocks connected by a triple point below the wedge corner. Also, both these cases result in vortex dipole similar to that of $\theta_{2w} = 60°$ and direction of the movement of dipole vortex, too, are similar. Thus, these shock interactions with the vortex present at the wedge corner result in highly complex flow structures.

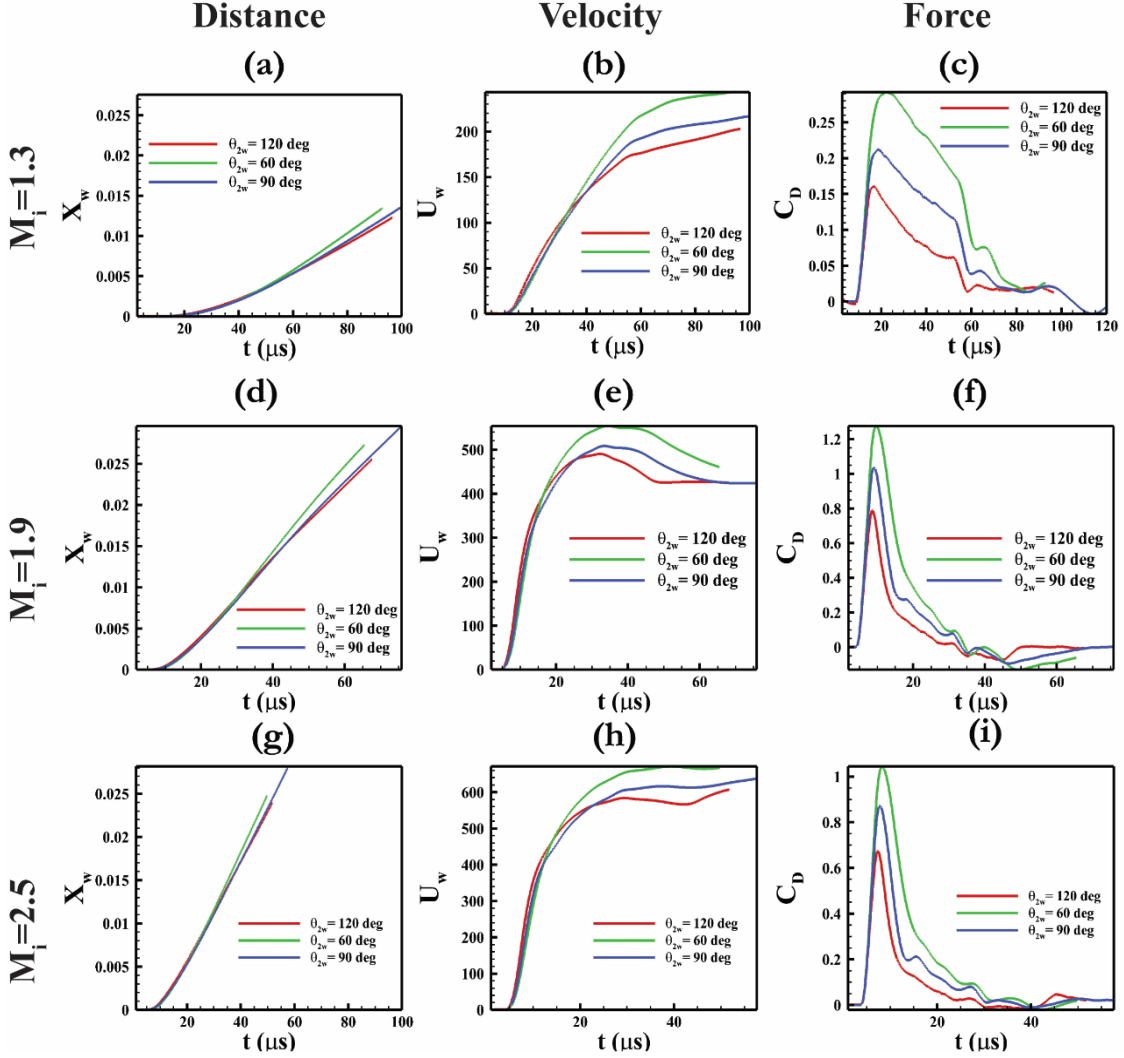

**Figure 16:** Time evolution plots. First column: Distance $X_w$ (a, d, g); Second Column: Velocity $U_w$ (b, e, h); Third Column: Drag coefficient $C_D$ of the shock driven wedges (c, f, i)

**Fig.16** depicts the time history of the distance $X_w$, velocity $U_w$, and drag force $C_D$ for experienced by the wedges for all the cases (see Appendix A for the description of how $X_w$, $U_w$ and $C_D$ are calculated). The plots show that for all incident Mach numbers $\theta_{2w} = 60°$ case has a steeper slope for distance followed by $\theta_{2w} = 90°$ and $120°$. The velocity magnitude too follows the same trend.

Before discussing the drag force plot, it is noteworthy to mention that force is calculated by integrating pressure and shear stress experienced by the wedge both due to the shock impact on them and as well as due to the wedge acceleration. As the shock passes through the wedge, the drag force-time history shows an early initial peak followed by drag dropping to zero levels as the shock moves beyond the wedge. By looking at shock locations at the time instance when the wedge is experiencing maximum drag (see **Fig.17**), it can be inferred that the peak is mostly due to the enormous pressure difference the wedge experiences as the shock initially passes through the wedge. As the shock passes through the wedge, this pressure difference too progressively gets reduced, resulting in a reduction of drag force. This drag curve trend is observed in all cases. The drag plots in **Fig.16** shows that at $M_i = 1.3$, the $C_D$ has a broader peak curve for all corner angles. For $M_i = 1.9$ and 2.5 it is narrow.

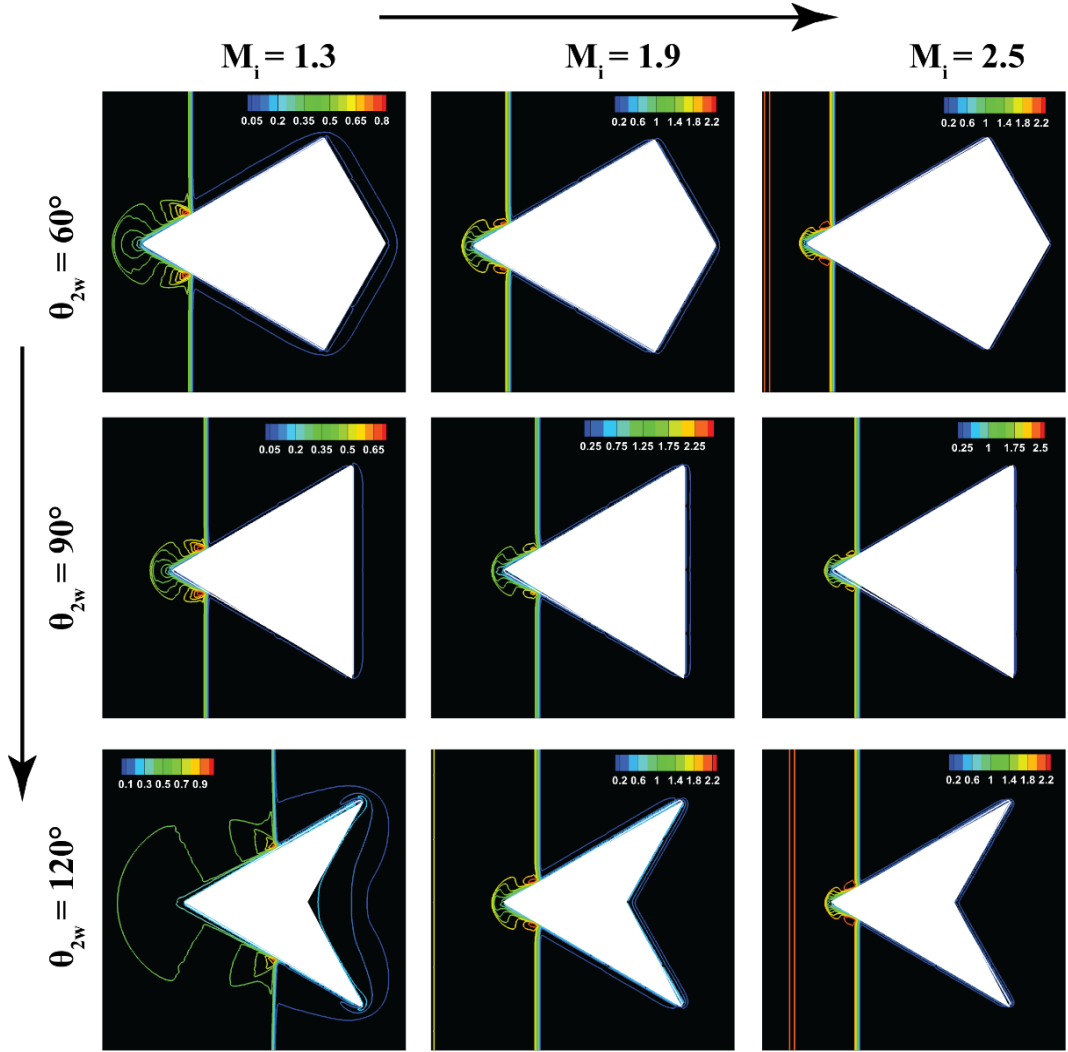

**Figure 17:** Mach Contour plot corresponding to the time instances when the drag force is at peak.

The drag plots in **Fig.16** exhibit that $C_{Dmax}$ increases at first when Mach number is increased from $M_i$ = 1.3 to 1.9 but decreases on further increasing the Mach number to $M_i$ = 2.5. This trend reversal is due to the following reasons. **Fig.18** depicts pressure co-efficient contour plots corresponding to time instance when the wedges had reached peak drag force. Cases corresponding to $M_i$ = 1.3 show the wedge experience lower pressure difference, while for $M_i$ = 1.9 and 2.5, the pressure difference is higher compared to the $M_i$ = 1.3. Note that $M_i$ = 2.5 cases experience lower pressure difference than $M_i$ = 1.9. This is because with increasing Mach number, the relative motion between the moving shock and moving wedge results in moving wedge witnessing reduced effective Mach number, which results in lower drag force than the $M_i$ = 1.9 cases. This has been observed by Luo et al.[44] in the case of shock driven cylinder cases. **Fig.16** drag plots also depict that $\theta_{2w}$ = 60° has higher $C_{Dmax}$ than the $\theta_{2w}$ = 90° and 120° for all the cases. Increasing the corner angle has the effect of increasing the pressure behind the wedge, which reduces the overall pressure difference resulting in drag reduction. This is much more evident when we look at the pressure distribution plot from **Fig.18** corresponding to $M_i$ = 1.3

The time evolution of the triple point (TP) trajectory for the three corner angles at different Mach numbers is presented in **Fig.19.** The first column shows the time evolution of triple point in x-direction while the second column shows of y-direction. Both stationary and moving cases are presented for comparison purposes. All the trajectories for the moving cases are calculated relative to the fixed origin. For stationary cases, as the Mach number increases, the amplitude of TP increases both in x and y-direction. The amplitude of moving cases show that their amplitude is higher than the stationary cases. But the amplitude increase follows a similar trend as the drag force trend. The amplitude increases with an increase in Mach number from $M_i$ = 1.3 to 1.9 but decreases when the Mach number further increases from $M_i$ = 1.9 to 2.5. As pointed out earlier, this trend reversal is because of the relative motion between moving shock and moving wedge.

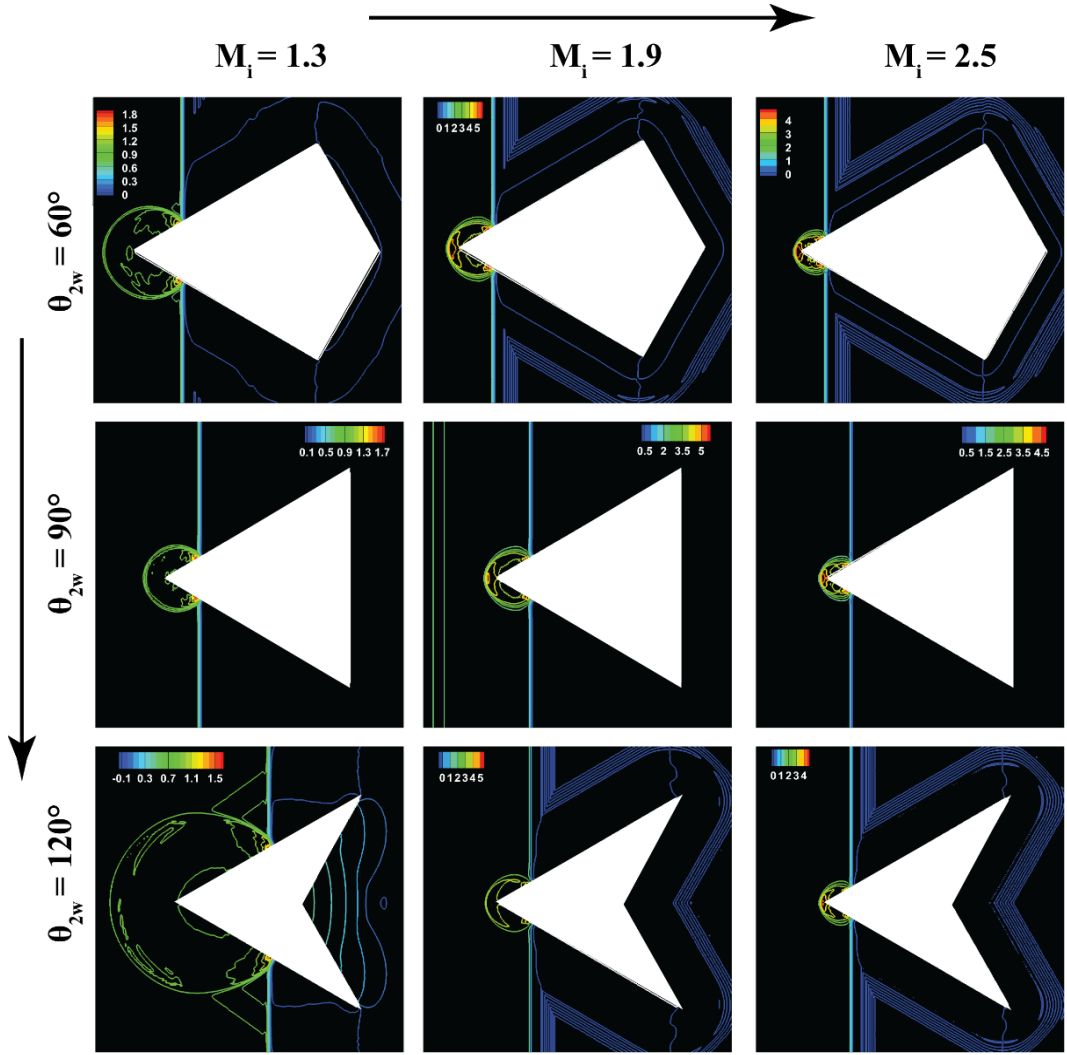

**Figure 18:** Pressure Contour plot corresponding to the time instances when the drag force is at peak.

**Fig.20** presents the numerical schlieren image of all the moving cases in the parametric space considered ($M_i, \theta_{2w}$). The flow structures are symmetric between upper and lower halves of the wedge, similar to stationary cases. Irrespective of the wall angle, at $M_i$ = 1.3, the reflected shock follows the wave configuration of Single Mach Reflection (SMR). At higher incident Mach numbers $M_i$ = 1.9 to 2.5, the reflected shock transitions into Double Mach Reflection (DMR). Also, the boundary layer separation at the first wall ($\theta_{1w}$) and formation of jets at the front tip of the wedge are observed clearly at higher incident Mach numbers. Changing the second wall angle does not make any impact on the transitions. But the shock vortex interactions near the wedge corners are influenced by the shock structure that forms behind the wedge as it accelerates forward.

It is worth mentioning here that so far to study these transition in shock wave reflection configurations, unsteady shock wave reflections are usually obtained in one of the three ways[4]: Reflecting a shock wave with a constant velocity over a non-straight surface or on a straight surface with a shock wave of non-constant velocity or using a shock wave with a non-constant velocity over the non-straight surface. Through this study, we have demonstrated that shock driven finite wedge problems could also be used to generate these shock wave transition phenomena, which can shed more light on the phenomena in more realistic settings.

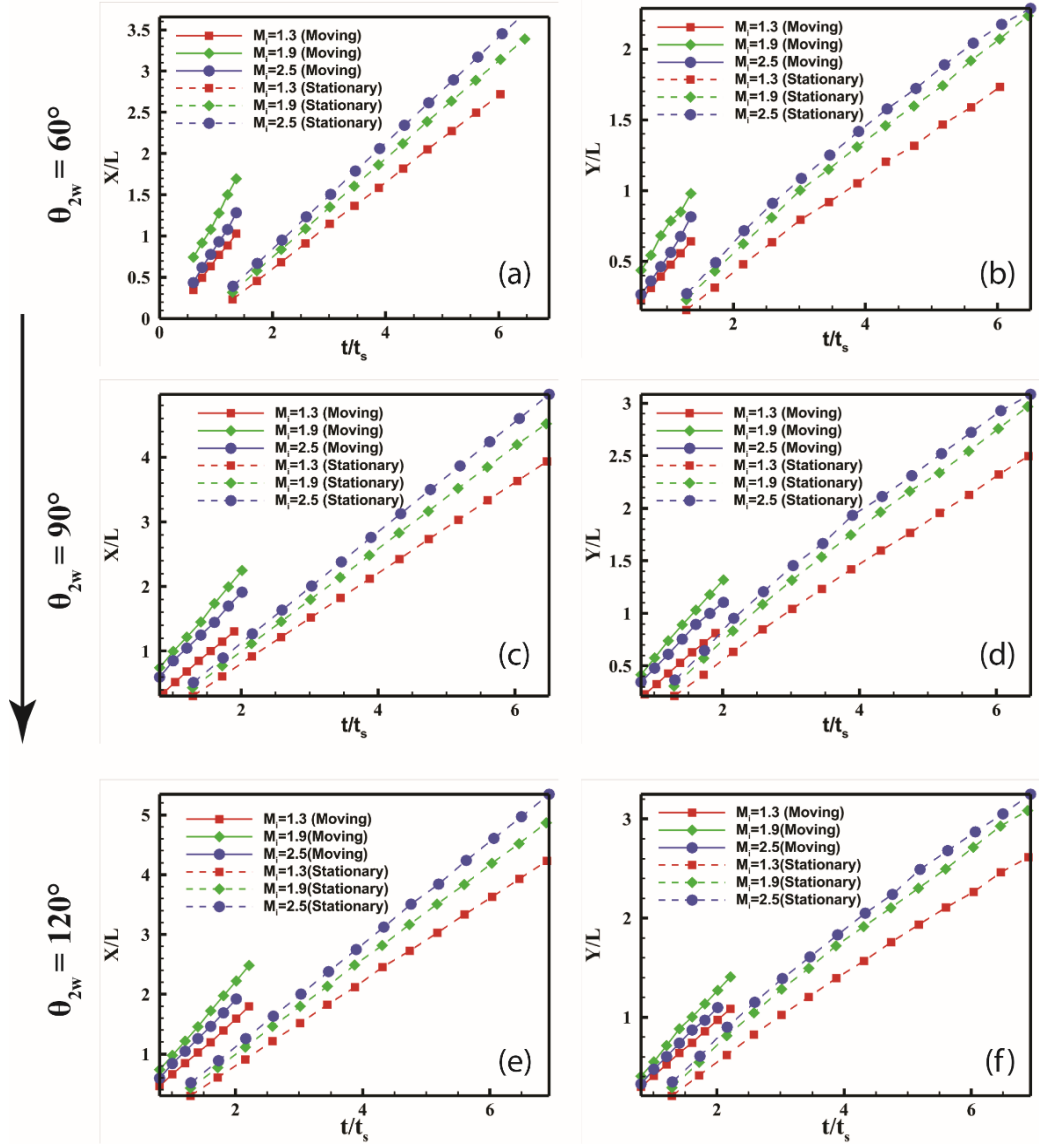

**Figure 19:** Triple Point Trajectory for moving cases compared to stationary cases (a-b) $\theta_{2w} = 60°$ (c-d) $\theta_{2w} = 90°$ (e-f) $\theta_{2w} = 120°$

## IV. Conclusion

The present article numerically investigates the shock wave interaction with a stationary wedge and moving wedge. A sharp interface immersed boundary approach in conjunction with a fifth-order WENO scheme is used to simulate the flow field. The focus remains on analyzing the impact of the incident Mach number and the second wall angle on the flow structure.

Both the incident Mach number ($M_i$) and second wall angle ($\theta_{2w}$) have a strong effect on the interaction process. In stationary cases, the observation suggests that a smaller wall angle hinders the growth of the vortex. Beyond $\theta_{2w} = 90°$, the rate of circulation production remains constant. Upon analyzing the different terms in the vorticity transport equation, the involved vortex dynamics reveal that vorticity diffusion due to viscous effects is dominating in vorticity production.

Finally, the shock driven wedge cases have been simulated to study the transition of reflected shock wave configuration as the wedge accelerates forward due to moving shock. It is found that at lower Mach numbers, no transition takes place. But at higher Mach number, the reflected shock transitioned from Single Mach Reflection to Double Mach Reflection. The intermediary state of Transition Mach Reflection (TMR) was also observed. The time history of drag coefficient shows that the maximum drag coefficient is found to be in case of $\theta_{2w} = 60°$ for all incident Mach numbers.

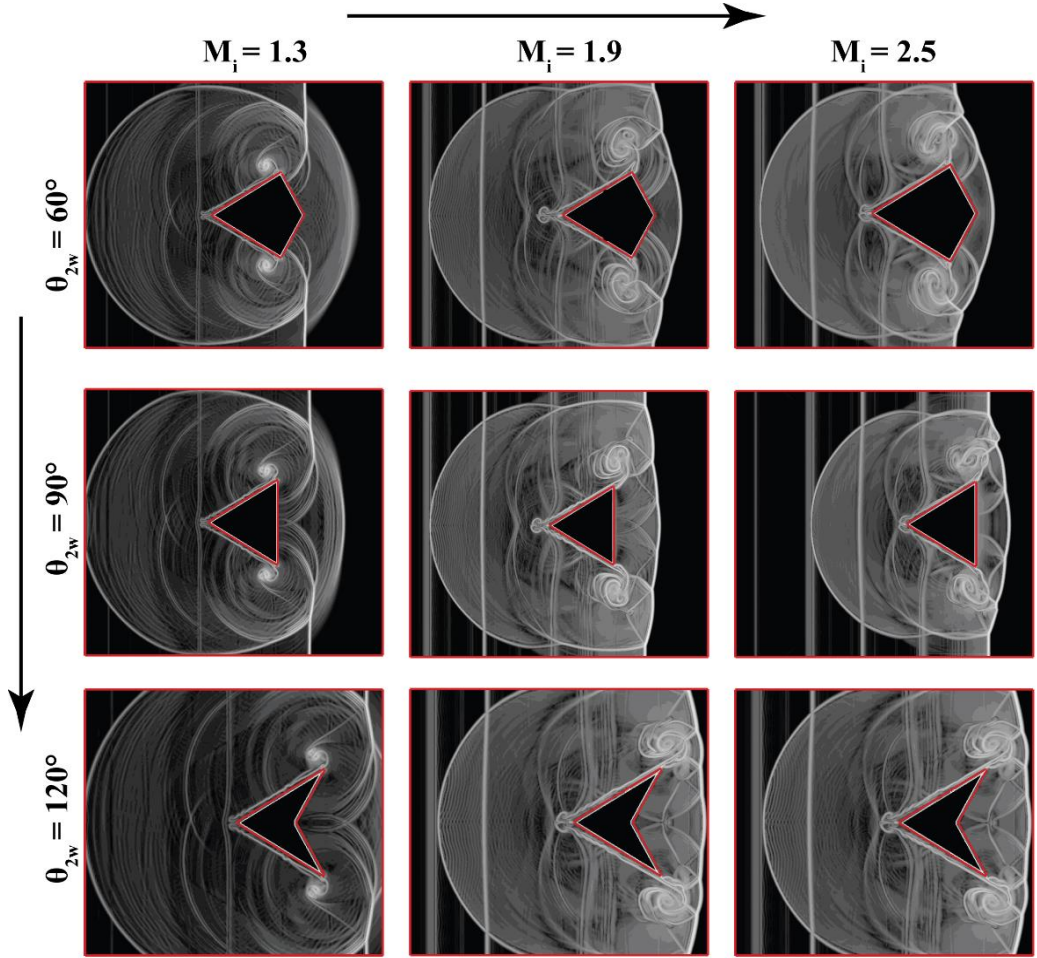

**Figure 20:** Array of numerical Schlieren images showing flow structures resulting from complex interactions of moving shock-driven wedge cases. Left to right shows the variation of flow structure with increasing incident shock Mach number while top to bottom shows the flow structure variation with an increase in wall angle. The three columns $M_i$ =1.3, 1.9, 2.5 correspond to the physical times t = 73.5μs, 51.84μs, 38.9μs respectively

## Appendix A: Details about Numerical Methods

As mentioned in the section.IIB, we use sharp interface immersed boundary framework[53-57] which reconstructs the flow field variables of nodes that are in the immediate vicinity of the immersed body (Wedge (w) in this study) such that it satisfies three important boundary conditions at the interface namely

(i)     Dirichlet Velocity Boundary condition:

$$U_f(t) = \begin{cases} U_w(t) & Moving\,body \\ 0 & Stationary\,body \end{cases} \quad (A1)$$

Where $U_f$ is fluid velocity vector, $U_w$ is the velocity of the wedge.

(ii)     Neumann Pressure Boundary condition:

$$-\left(\frac{\partial p}{\partial n}\right)_w = \frac{\rho_s U_{ft}^2}{R} - \rho_s a_n \quad (A2)$$

where $U_{ft}$ is the tangential component of the velocity field at the immersed interface. $a_n$ is the normal component of the acceleration of the interface.

(iii)     The heat flux along the normal direction to the immersed boundary surface was assumed to be zero in this article.

$$\left(\frac{\partial T}{\partial n}\right)_w = 0 \quad (A3)$$

The forces on the immersed body are calculated by integrating the pressure and shear stress at the wedge surface through the following expression.

$$F_i = \int_w [-P\delta_{ij} + \tau_{ij}] n_j dw \quad (A4)$$

Moving body cases presented in this study, where shock interaction with wedge results in rigid motions, the forces obtained through the above procedure is utilized to update the information regarding velocity and position of the solid at the advanced time levels. Verlet integration approach is used for the purpose.

$$U_{i,w}^{n+1} = U_{i,w}^n + \frac{\Delta t}{M_w} F_i^n$$

$$X_{i,w}^{n+1} = X_{i,w}^n + U_{i,w}^n \Delta t + \frac{(\Delta t)^2}{M_w} F_i^n \quad (A5)$$

Where $U_{i,w}$ is the velocity vector of wedge and $X_{i,w}$ position vector in i[th] Cartesian coordinate direction. Further information about the framework can be found in our recent publications.

The Drag coefficient $C_D$ is calculated from the expression,

$$C_D = \frac{F_x}{0.5\rho U_s^2} \quad (A6)$$

Where $U_s$ is the velocity of the incident shock.

## Data Availability

The data that support the findings of this study are available from the corresponding author upon reasonable request.

## Acknowledgments

Simulations are carried out on the computers provided by the Indian Institute of Technology Kanpur (IITK) (www.iitk.ac.in/cc), and the manuscript preparation, as well as data analysis, has been carried out using the resources available at IITK. This support is gratefully acknowledged.